\documentclass[12pt,preprint,eqsecnum]{aastex}
\usepackage{psfig}

\newcommand{\et}{{\rm et al.}}
\def\earth{\hbox{$\oplus$}}

\newfont{\boldfont}{cmmib10}

\newcommand{\btheta}{\hbox{\boldfont \symbol{18} }}

\begin{document}
\title{Microarcsecond Radio Imaging using Earth Orbit Synthesis}
\author{Jean-Pierre Macquart\altaffilmark{1} \\
Research Centre for Theoretical Astrophysics,\\
School of Physics, University of Sydney, NSW 2006, Australia.\\
and Kapteyn Institute, University of Groningen, \\
Postbus 800, 9700 AV Groningen, The Netherlands\\
email: jpm@astro.rug.nl}
\author{David L. Jauncey \\
Australia Telescope National Facility, \\ 
CSIRO, PO Box 76, Epping NSW 1710, Australia\\
email David.Jauncey@atnf.csiro.au}

\begin{abstract}
The observed interstellar scintillation pattern of an intra-day variable 
radio source is influenced by its source structure.  If the velocity of the 
interstellar medium responsible for the scattering is comparable to 
the earth's, the vector sum of these allows an observer to probe 
the scintillation pattern of a source in two dimensions and, in turn, 
to probe two-dimensional source structure on scales 
comparable to the angular scale of the scintillation pattern, 
typically $\sim 10\,\mu$as for weak scattering.  We review the theory on the extraction of 
an ``image'' from the scintillation properties of a source, and show 
how earth's orbital motion changes a source's observed scintillation properties 
during the course of a year.  The imaging process, which we call 
Earth Orbit Synthesis, requires 
measurements of the statistical properties of the scintillations at 
epochs spread throughout the course of a year.
\end{abstract}


\section{Introduction}

Flux density variability on intra-day timescales has recently been  detected in
a few tens of extragalactic radio sources (Heeschen 1984;  Witzel \et\ 1986,
Kedziora-Chudczer \et\ 2001).  There is  now
overwhelming evidence that the principal mechanism responsible for intra-day
variability (IDV) at centimeter wavelength is interstellar scintillation (ISS),
rather than intrinsic source variability (Kedziora-Chudczer \et\  1997;
Dennett-Thorpe \& de Bruyn 2001a,b; Jauncey \et\ 2001; Rickett \et\  2001;
Jauncey \& Macquart 2001).

ISS is observed in compact ($\lesssim 100\,\mu$as) radio sources due to the 
scattering of radiation in the Galactic interstellar medium (ISM) as it 
propagates toward earth.  Inhomogeneities in the ISM impart phase structure  on
the wavefront, which leads in turn to a pattern of flux density deviations on
the observer's plane.   Flux density variability is observed in a scattered
source if an  observer moves relative to the source's scintillation pattern,
ie. due to  relative motion between the ISM, the source and the earth.  The
presence of such radio IDV implies extremely small source sizes ($\sim
5-100\,\mu$as), and high brightness temperatures ($T_B \sim 10^{13}-10^{15}$~K,
e.g. Kedziora-Chudczer \et\ 1997). This  is well above the inverse 
Compton limit, and it is not clear whether such emission is  explainable in
terms of Doppler  boosted synchrotron emission (Begelman, Rees \& Sikora
1994).  The  properties --- including the radio structures --- of IDV sources
are  therefore of considerable interest.

The scintillation pattern of an IDV source contains information on the 
structure of the source, which can be derived from   the
statistical properties of the scintillation pattern given the  statistical
properties of the scattering medium.   This has already been demonstrated for
sources undergoing interplanetary scintillation;  the experiment by
Cornwell,  Anantharamaiah \& Narayan (1989) is an elegant
illustration of this technique.

Some IDV sources exhibit large annual modulations in their  variability
timescales (Dennett-Thorpe \& de Bruyn 2001a;  Rickett \et\ 2001; Jauncey \&
Macquart 2001).  This is explained 
in terms of the earth's orbital motion; as the earth's velocity changes during
its orbit about the sun, so does an earth-based  observer's velocity relative to
the ISM. Both the speed and direction  change greatly if the speed of the
scintillation pattern relative to  the heliocentre, due to the intrinsic motion
of the ISM, is comparable  to the earth's speed.  This effect is expected to be
prevalent for IDV  sources because the ISM velocity is comparable to earth's,
but is not generally observed in the scintillations of pulsars because of
their high intrinsic speeds.  Their speeds, typically $\gtrsim  150$~km/s are
much greater than that  of the earth or the ISM, and usually dominate the
scintillation speed.

The large variation in scintillation speed and direction over the 
course of a year makes it possible to observe two-dimensional 
structure in the scintillation pattern.  Hence ISS can be used to ``image'' a 
source on $5-100\,\mu$as scales.  In this paper we show how to use changes in 
the earth's velocity around the sun to probe the scintillation pattern 
in two dimensions, and how to invert this information to obtain source 
structure.  We refer to this technique as {\it Earth Orbit Synthesis} 
in light of its similarity to the conventional radio interferometric 
imaging technique of Earth Rotation Synthesis.

As for the radio imaging technique of Earth Rotation Synthesis, Earth
Orbit Synthesis assumes that the source structure remains relatively
stable over the course of the observations.  This appears to be a good
assumption for several IDV sources that have been well-studied over a number
of years.  B0917$+$624 appears to have been scintillating over the decade and a
half that this source has been followed in detail (e.g. Jauncey \& Macquart
2001), while J1819$+$3845 has shown a clear annual cycle that repeats closely
over at least the last two years (Dennett-Thorpe \& de Bruyn 2001a), and
PKS~1519$-$273 has shown IDV each time it has been observed in sufficient
detail over the past decade (Kedziora-Chudczer \et\ 2001).  However, it should
be remembered that this technique is not applicable to those sources that
exhibit episodic IDV such as PKS~0405$-$385 (Kedziora-Chudczer \et\ 2000).



In \S\ref{Emotion} the apparent scintillation velocity of an 
extragalactic source is derived in terms of the velocities of the ISM 
and the earth.  In \S\ref{SrcStruct} we relate the statistics of 
intensity and polarization fluctuations to source structure.  A model 
for the scintillation velocity is combined with statistics of the 
intensity fluctuations to derive source structure in \S\ref{Combined}, 
and the technique is demonstrated with some specific examples.  The 
conclusions are presented in \S\ref{Concl}.

\section{Earth's orbital motion} \label{Emotion}

The velocity of an earth-based observer across the 
scintillation pattern of a scattered source varies annually due 
to the earth's orbit about the sun.  Here we derive the observed
scintillation velocity in terms of the position of a distant 
scintillating source and the velocity of the interstellar medium 
relative to the heliocentre (see Fig. 1).

We assume that the internal motions of the scattering material  are negligible
compared to the bulk motion of the material across the  line of sight.  We also
assume that the scattering occurs  predominantly in one layer of the ISM only,
at a distance $L$ from earth, so that motion of the  scattering material can be
characterized by a single velocity. This assumption appears to be valid for a
few IDV radio sources whose scattering screens appear to be nearby and
localized (e.g. Dennett-Thorpe \& de Bruyn 2000),
and is a good approximation for other IDV sources.

The observed velocity of the line of sight to an extragalactic source 
through a scattering screen is given by
\begin{eqnarray}
{\bf v}_{\rm obs} = {\bf v}_{\rm ISM} - {\bf v}_{\earth},
\end{eqnarray} 
where ${\bf v}_{\rm ISM}=({v_{\rm ISM}}_x,{v_{\rm ISM}}_y,{v_{\rm 
ISM}}_z)$ and ${\bf v}_{\earth}$ are, respectively, the velocities of 
the ISM and the earth relative to the heliocentre.  For extragalactic sources,
the proper motions of the source across the line of sight can be ignored.  The co-ordinate system is aligned so that 
${\bf e}_z$ points normal to the ecliptic plane and ${\bf e}_x$ points 
along ecliptic longitude $0^{\circ}$ in the ecliptic plane.

We ignore the $0.01^{\circ}$ inclination of the earth's orbit to the 
ecliptic, and approximate the earth's orbital motion as occurring in 
the ecliptic plane.  The eccentricity of the earth's orbit 
($e=0.0167$) results in a negligible variation in orbital speed.  We 
therefore approximate the earth's orbital velocity relative to the 
heliocentre as:
\begin{eqnarray}
{\bf v}_{\earth} &=& v_{\earth} (\sin 2 \pi t, -\cos 2 \pi t,0),
\end{eqnarray}
where $v_{\earth}=29.78\,$km/s is the mean orbital speed of the earth, 
$t$ is the time in years measured from the vernal equinox.

As shown in Figure 1, and choosing the earth to be at the origin of the 
co-ordinate system we 
write location of the scintillation screen in terms of ecliptic co-ordinates 
${\bf a}=a\,(\cos \phi \cos\theta,\sin \phi \cos \theta,\sin \theta)$, 
where $\phi$ and $\theta$ are, respectively, the ecliptic longitude 
and latitude of the scintillating source.  For typical screen distances
$\gtrsim 20$pc, changes in $\hat{\bf a}={\bf a}/a$ caused by orbital 
changes in the earth's position relative to the screen are negligible.

The observed motion of the scintillation pattern across the line of 
sight is
\begin{eqnarray}
{\bf v}_{\rm scint} = [V_\parallel + v_{\earth} \cos(2 \pi t - \phi) ] 
{\bf e}_\parallel + [V_\perp + v_{\earth} \sin \theta \sin (2 \pi t-\phi)] 
{\bf e}_\perp, \label{ScintVelocity}
\end{eqnarray}
where ${\bf e}_\parallel = (-\sin \phi,\cos\phi,0)$ and ${\bf e}_\perp 
=(-\cos \phi \sin \theta, -\sin \phi \sin \theta,\cos \theta)$ are, 
respectively, the unit vectors parallel and perpendicular to the 
ecliptic along the line of sight to the scintillating source.  The 
intrinsic speed of the ISM relative to the heliocentre and tangential 
to the line of sight is $V_{\parallel}$ along ${\bf e}_{\parallel}$ 
and $V_\perp$ along ${\bf e}_\perp$, where
\begin{eqnarray}
V_\parallel &=& {v_{\rm ISM}}_y \cos \phi - {v_{\rm ISM}}_x \sin \phi, \\
V_\perp     &=& {v_{\rm ISM}}_z \cos \theta - 
\sin \theta ({v_{\rm ISM}}_x \cos \phi + {v_{\rm ISM}}_y \sin \phi).
\end{eqnarray}

Equation (\ref{ScintVelocity}) implies that the velocity of the scintillation
pattern relative to the earth changes in both magnitude and direction on an
annual cycle.

\section{Source structure} \label{SrcStruct}
The scintillation pattern of a scattered source is related to both its 
intrinsic structure and the inhomogeneities in the ISM. Source 
structure can be derived if the statistical properties of the 
scintillation pattern due to a point-source are known.  In 
\S\ref{Point} we express for the scintillation pattern due to a point 
source in terms of the scattering medium's statistical properties.  In 
\S\ref{Extend} we show how the scintillation pattern is modified when 
the scintillating source has extended structure.

\subsection{The scintillation pattern of a point source} 
\label{Point}
The scintillation pattern of a point source depends upon the phase 
fluctuations on the scattering screen which are characterized by the 
phase structure function
\begin{eqnarray}
D_{\phi}({\bf r}) = \langle [\phi({\bf r}'+{\bf r}) - 
\phi({\bf r}')]^2 \rangle .\label{Dphi}
\end{eqnarray}
The angular brackets denote an ensemble average over all possible 
realisations of the scattering medium, and ${\bf r}$ is a two-dimensional vector
on the scattering screen.

For most lines of sight through the ISM the phase structure function 
reflects an underlying power law spectrum of phase inhomogeneities 
between some inner and outer scales $l_0$ and $L_0$ respectively, and 
one has
\begin{eqnarray}
    D_{\phi} ({\bf r}) = \left( \frac{r}{r_{\rm diff}} 
    \right)^{\beta-2}, \label{DphiPwr}
\end{eqnarray}
where $\beta$ is close to $11/3$ (Armstrong, Rickett \& Spangler 
1995), the value expected for Kolmogorov turbulence.  The strength of 
the phase fluctuations is measured by $r_{\rm diff}$, the length scale 
over which the root-mean-square phase difference on the phase screen 
is one radian.

The strength of the scattering is determined by the ratio $r_{\rm 
F}/r_{\rm diff}$; the quantity $r_{\rm F}=\sqrt{c L/(2 \nu \pi)}$ is 
the Fresnel scale.  The scattering is weak when phase irregularities on the 
scattering screen are small, when $r_{\rm diff} > r_{\rm F}$ and 
strong when they are large, when $r_{\rm diff} < r_{\rm F}$.  The 
scattering strength varies as a function of frequency since both 
$r_{\rm diff} \propto \nu^{2/(\beta-2)}$ and $r_{\rm F}$ vary with 
frequency.  It is convenient to define $\nu_t$ as the frequency at 
which the scattering changes from strong to weak.  The 
scattering is weak at frequencies above $\nu_t$, while strong 
scattering occurs at frequencies below $\nu_t$.

The qualitative properties of the scintillation depend heavily on the 
scattering strength, and hence on $\nu_t$.  In general, the value of 
$\nu_t$ toward a scintillating source is unknown, but can be 
determined from observations of its variability over a range of 
frequencies (e.g. Kedziora-Chudczer \et\ 1997; Quirrenbach \et\ 2000; 
Kedziora-Chudczer \et\ 2001), 
or can be inferred from models of the ISM (Walker 1998).  
Along lines of sight out of the Galactic plane the transition frequency 
is in the range $\nu_t \sim 3-8$~GHz (Walker 1998).  

A useful statistic of the intensity fluctuations of a point source due to scattering by the 
ISM is the intensity autocorrelation function
\begin{eqnarray}
{C_{II}}_{\rm pt} ({\bf r}) = \langle \delta I_{\rm pt}({\bf r}'+{\bf 
r})  \delta I_{\rm pt}({\bf r}') \rangle,
\end{eqnarray}
where $\delta I_{\rm pt}({\bf r}) = 
[I({\bf r}) - \langle I \rangle]/\langle I \rangle$ is the 
fractional deviation in the intensity at the point ${\bf r}$ on the 
observer's plane.  This function describes the 
correlation between the intensity fluctuations between points ${\bf r}'$ 
and ${\bf r}'+{\bf r}$ on the observer's plane.  An alternative but equivalent 
representation of the 
intensity fluctuations is provided by the power spectrum, 
${W_{II}}_{\rm pt}$, which is 
related to the intensity autocorrelation function by a Fourier transform
\begin{eqnarray}
{W_{II}}_{\rm pt} = \int d^2 {\bf r} \exp[i {\bf q} \cdot {\bf r}] 
{C_{II}}_{\rm pt} ({\bf r}). \label{CWrelation}
\end{eqnarray}
We adopt the power spectrum representation in the discussion below.

The functional dependence of ${W_{II}}_{\rm pt} ({\bf r})$ depends on 
the scattering strength.  In the regime of weak scattering, one has 
(Salpeter 1967, Jokipii \& Hollweg 1970, Narayan 1992) 
\begin{eqnarray}
{W_{II}}_{\rm pt} ({\bf q}) &=&  C \, q^{-2} \left(
\frac{1}{q r_{\rm diff}} \right)^{\beta-2}  \sin^2 
\left(\frac{1}{2} r_{\rm F}^2 q^2 \right), \label{WeakWII}
\end{eqnarray}
with $C=11.2$ for $\beta=11/3$.
The power spectrum peaks at $q \sim 1/r_{\rm F}$, showing that the 
typical length scale of scintillations in the regime of weak 
scattering is $r \approx r_{\rm F}$.  An object undergoing weak 
scintillations would exhibit flux density variations on a timescale $t_{\rm F} 
= r_{\rm F}/v_{\rm scint}$.

In the regime of strong scattering the power spectrum of intensity fluctuations 
for $\beta<4$ is (Gochelashvily \& Shishov 1975) 
\begin{eqnarray}
{W_{II}}_{\rm pt} (q) = \left\{ \begin{array}{ll}
2^{\beta-3} 
\frac{\pi (\beta-2) (4-\beta) \Gamma(\beta/2)}{\Gamma\left( \frac{6-\beta}{2} \right)} 
\, q^{-2}  \left(\frac{r_{\rm F}}{r_{\rm diff}} \right)^{\beta-2} (q r_{\rm F})^{6-\beta} 
\exp[-(q r_{\rm ref})^{\beta-2}], &q \lesssim r_{\rm ref}^{-1} \\
\frac{2 \pi}{\beta-2} \, \Gamma \left(\frac{1}{\beta-2} \right) r_{\rm 
diff}^2, & r_{\rm ref}^{-1} \lesssim q \lesssim 
r_{\rm diff}^{-1} \\
\frac{2^{\beta} \pi (\beta-2) \Gamma(\beta/2)}{\Gamma \left( 
\frac{4-\beta}{2} \right)} \, q^{-2} (q r_{\rm diff})^{2-\beta}, & 
r_{\rm diff}^{-1} \lesssim q \\
\end{array}
\right. , \label{StrongWII}
\end{eqnarray}
where $r_{\rm ref}=r_{\rm F}^2/r_{\rm diff}$ is known as the 
refractive scale.
Intensity fluctuations can occur predominantly on two timescales in this 
regime, corresponding to the two peaks in ${W_{II}}_{\rm pt} (q)$.  
The outermost peak, corresponding to the effect of diffractive 
scintillation, occurs at $q \sim 1/r_{\rm diff}$ and is associated with 
fast-timescale, narrowband, large amplitude (up to 100\% modulated) variability.  

The other peak in ${W_{II}}_{\rm pt} (q)$ occurs at $q \sim 1/r_{\rm 
ref}$, and is associated with refractive scintillation.
Refractive scintillations occur on a longer timescale 
than weak scintillations, are broadband and have a typically low 
($\sim 10$\%) modulation index.  The timescale associated 
with refractive scintillation is $t_{\rm ref} = r_{\rm ref}/v_{\rm 
scint}$.  
This is a factor $r_{\rm F}/r_{\rm diff}>1$ slower than for weak scattering.

Intensity fluctuations of IDV sources scintillating in the strong 
scattering regime are seen to occur on timescales $\sim 1-10$~days with amplitudes 
up to $\sim 10$\% of the total source intensity 
(e.g. Kedziora-Chudczer \et\ 2001).  This is attributed to 
refractive scintillation which is exhibited by any source with 
angular diameter $\theta_S \lesssim r_{\rm ref}/L$.  No IDV source has 
yet been observed to exhibit diffractive 
scintillation.  The stringent limitations on source size required for 
a source to exhibit diffractive scintillation, $\theta_S \lesssim 
r_{\rm diff}/L$ are harder to satisfy for IDV sources.  For 
example, one requires $\theta_S \lesssim 6\,\mu$as at $\nu=2$~GHz to exhibit 
fully modulated diffractive scintillation scattering, where we assume 
$\nu_t=5$~GHz and $D=100$~pc.  These limitations are less stringent if
$D$ is smaller, as has been suggested for at least one source (Dennett-Thorpe \& de Bruyn
2000).

Expressions (\ref{WeakWII}) and (\ref{StrongWII}) are only valid when 
the statistical properties of the phase fluctuations on the scattering 
screen are isotropic, in the sense that the length scale $r_{\rm 
diff}$ does not vary with orientation on the screen.  Anisotropy 
introduces a preferred direction in the scattering medium so that 
${C_{II}}_{\rm pt}$ is no longer a function of $r$ only, but depends 
on the orientation.  Typical elongations of the scattering pattern in 
the ISM can exceed 2:1 (see Chandran \& Backer submitted for a full 
discussion).  Anisotropic screen structure can therefore cause elongations in the 
scintillation pattern similar to apparent elongations caused by intrinsic 
source structure.  One way to distinguish between source 
structure and screen anisotropy involves imaging the scattered source at low 
frequency, where the scattering is strong (ie.  $\nu < \nu_t$) and the 
condition $\theta_S < r_{\rm ref}/D$ is more likely to be satisfied.  
Then the shape of the scattered image is entirely dominated by the 
scattering and any anisotropy is readily apparent as an elongation of 
the scattering disk along the axis corresponding to small $r_{\rm 
diff}$.  This permits a direct measurement of the effects of screen 
anisotropy.  Of course, the measurement can only be attempted at frequencies 
low enough that the scattering disk is sufficiently large to be resolved; this 
may be an important limitation in practice.

We see in \S\ref{IntensityFlucts} 
that the effect of screen anisotropy is not important when the angular 
size of a source exceeds the angular scale of the scintillation 
pattern for a point source (ie.  $r_{\rm F}/D$ for weak 
scattering), and is readily distinguished from source structure by its 
characteristic structure in ${W_{II}}_{\rm pt}$.  
Thus, to first order, effects of screen anisotropy are diminished in many IDV 
sources.  Hereafter we restrict our discussion to isotropic phase screens for the purposes of 
simplicity.

It possible to determine which of equations (\ref{WeakWII}) and 
(\ref{StrongWII}) is applicable to a set of observations provided that 
the characteristics of the intensity fluctuations are known at several 
frequencies.  For weak scattering one has $t_{\rm F} \propto 
\nu^{1/2}$, whereas $t_{\rm ref} \propto \nu^{2.2}$ (assuming $\beta=11/3$) for 
strong scattering.  The large difference in the timescales of the intensity
variations allows a determination of whether the variability is occurring 
in the regime of weak or strong scattering.

Many sources are observed to scintillate at frequencies that straddle  the
regime  between weak and strong scattering.  It is difficult to provide 
analytic expressions for ${W_{II}}_{\rm pt}$ in this transition regime, and
instead, equation (\ref{WeakWII}) may be used on the understanding  that it is
only approximately correct, or ${W_{II}}_{\rm pt}$ may be derived  numerically
from scattering simulations.

\subsection{The scintillation pattern of an extended source} \label{Extend}

We now discuss the effect of source structure on the observed 
scintillation pattern of an extended source.  Direct measurements of the 
scattered {\em wavefield} from an extended source may be inverted to 
completely determine the source structure.  The {\em intensity} 
scintillation pattern of a source also contains information on source 
structure.  However, analysis of the intensity fluctuations alone only provides 
partial information on source brightness distribution.

\subsubsection{The wavefield}

The wavefield of a scattered source is directly related to its
structure.  The mutual coherence function is the 
simplest nontrivial moment of the wavefield.   If 
the wavefield measured by a receiver at the point ${\bf 
r}$ on the observer's plane is $u(z;{\bf r})$, the normalized two-point 
correlation function is then
\begin{eqnarray}
\Gamma_{\alpha \alpha'}(z;{\bf r})  = \frac{
\langle u_{\alpha}(z;{\bf r}') u_{\alpha'}^*(z;{\bf r}'+{\bf r}) 
\rangle}{I_0},
\end{eqnarray}
where $\alpha, \alpha' = [x,y]$ corresponds to the component of the 
wavefield oscillating along the $x$ and $y$ axes orthogonal to the 
direction of propagation, along the $z$-axis, and $I_0$ is the mean 
source total intensity. The normalized Stokes visibilities are defined by 
\begin{eqnarray}
\Gamma_I(z;{\bf r}) &=& (\Gamma_{xx}(z;{\bf r}) + \Gamma_{yy}(z;{\bf r}))/2 \qquad 
\Gamma_Q(z;{\bf r}) = (\Gamma_{xx}(z;{\bf r}) - \Gamma_{yy}(z;{\bf r}))/2 
  \nonumber \\ 
\Gamma_U(z;{\bf r}) &=& (\Gamma_{xy}(z;{\bf r}) + \Gamma_{yx}(z;{\bf r}))/2 \qquad
\Gamma_V(z;{\bf r}) = i (\Gamma_{xy}(z;{\bf r}) - \Gamma_{yx}(z;{\bf r}))/2.
\end{eqnarray}

It is assumed that the source is spatially incoherent and that the 
wavefront from the IDV source incident upon the scattering screen is 
planar, which is a very good approximation for extragalactic 
sources.  (The planar approximation is not valid for Galactic objects, but it 
is possible to make a simple correction for the curvature of the 
wavefront in this case (e.g., Goodman \& Narayan 1989).)  The 
mutual coherence function measured by an observer is then related to the mutual coherence function 
of the source $\Gamma_{X}(0;{\bf r}),\,X=[I,Q,U,V]$ by
\begin{eqnarray}
\left( \begin{array}{l}
 \Gamma_I (z;{\bf r}) \\
 \Gamma_Q (z;{\bf r}) \\
 \Gamma_U (z;{\bf r}) \\
 \Gamma_V (z;{\bf r})  
 \end{array}
\right) = 
e^{-D_{\phi} ({\bf r})/2} \left( \begin{array}{c}
\Gamma_I (0;{\bf r}) \\
\Gamma_Q (0;{\bf r}) \cos 2 \phi_V + \Gamma_U (0;{\bf r}) \sin 2 
\phi_V \\
- \Gamma_Q (0;{\bf r}) \sin 2 \phi_V + \Gamma_U (0;{\bf r}) \cos 2 
\phi_V \\
\Gamma_V (0;{\bf r})
\end{array}
\right), \label{MutCoher}
\end{eqnarray}
where $\phi_V = \lambda^2 {\rm RM}$ is the mean phase change induced by 
Faraday rotation along the ray path, and we assume no inhomogeneity in the
Faraday rotation transverse to the line of sight. 

The normalized mutual coherence of the source, $\Gamma_X(0;{\bf r})$ is related 
to its angular brightness distribution normalized by the mean intensity via a Fourier transform
\begin{eqnarray}
\Gamma_X(0;{\bf r})= \int d^2{\btheta} \, b_X(\btheta) \exp \left[-2 \pi i 
\frac{\btheta \cdot {\bf r}}{\lambda} \right]. \label{BrightnessDist}    
\end{eqnarray}
The behaviour of the mutual coherence 
function depends on the angular size of the source relative to the 
angular scale of the scintillation pattern as seen by an observer on 
the earth.  Equations (\ref{MutCoher}) and 
(\ref{BrightnessDist}) imply the behaviour of $\Gamma(z;{\bf r})$ is 
dominated by the source angular structure if its angular size exceeds 
that of the scintillation pattern seen by an observer, $\theta_S \gtrsim k / r_{\rm diff}$.  
In this instance the source is said to be resolved by the scattering 
medium.

Conversely, the mutual coherence function of a source with angular 
size $\theta_S \lesssim k/r_{\rm diff}$ is dominated by the 
scattering, and $\Gamma(z;{\bf r}) \sim \Gamma(z;0) \exp[-D_\phi ({\bf 
r})/2]$.  It is difficult to derive source structure when the source 
is unresolved by the scattering medium because the scattering 
properties dominate the mutual coherence function.  
This limitation is important in the regime of strong scattering where 
$k/r_{\rm diff}$ is may be large relative to the size of an IDV source.

The quantities $\Gamma_X({\bf r})$ can be measured over short coherent
integrations (see, e.g. Goodman \& Narayan 1989); typically two or three 
scintles are sufficient to obtain a good estimate.

\subsubsection{Intensity fluctuations}

Measurement of the intensity fluctuations contains partial information 
on the structure of a source.  In this section we derive the 
relationship between the brightness distribution of a scattered source 
and its intensity fluctuations.  The pattern of intensity fluctuations 
from an extended source is derived by dividing the source into small 
elements and computing the individual intensity scintillation pattern 
from each element (again assuming a spatially incoherent source).  The 
scintillation pattern of the extended source is therefore the sum of 
the scintillation patterns from all the elements of the source.

Figure 2 shows that the wavefront incident on the screen at an angle 
${\btheta}$ to the line of sight, and heading towards the point ${\bf 
r}+L{\btheta}$, intersects a {\it thin} scattering screen at the same point as a 
normally incident wavefront, heading towards ${\bf r}$.  Each of these 
wavefronts experiences identical scattering, and the amplification 
caused by the scattering screen is thus identical for these two 
wavefronts.  It follows that scintillation pattern of an extended source 
is (Little \& Hewish 1966),
\begin{eqnarray}
\delta S_X({\bf r}) = \int d^2 {\btheta}\, b_X ({\btheta}) 
\delta I_{\rm pt}({\bf r}-L {\btheta}),  \label{ExtendSrc}
\end{eqnarray}
where $\delta S_X = (S_X - \langle S_X \rangle)/\langle S_X \rangle$ is 
the fractional flux density deviation in Stokes parameter $X=[I,Q,U,V]$, and 
$b_X({\btheta})$ is the angular brightness distribution of the source 
as observed {\it at} the scattering screen, normalized such that $\int b_I({\btheta}) d^2{\btheta} = 1$ and $\int 
b_Y({\btheta}) d^2{\btheta} = Y/I_0, \,\, Y=[Q,U,V]$.
The brightness distributions $b_Q$ and $b_U$ at the screen differ from their 
distributions at the source if Galactic Faraday rotation is important. 
The distributions at the source are found by determining the Faraday 
rotation along the line of sight to the source.  A magnetized ISM may 
also induce {\it fluctuations} in the amount of radiation transferred 
between Stokes $Q$, $U$ and $V$, however the fluctuations in the 
degree of birefringence are of order $\Omega_e/\omega_p \sim 
10^{-8}$ smaller than the phase fluctuations considered here, and are 
probably unimportant for the scattering considered 
here (see Macquart \& Melrose 2000).

The flux density deviations observed in an extended source are 
connected to the scintillation statistics of a point source by 
\begin{eqnarray}
C_{XY}({\bf r}) &=& \langle \delta S_X({\bf r}') \delta S_Y({\bf r} + 
{\bf r}') \rangle \nonumber \\
&=& \int d^2 {\btheta} d^2 {\btheta}' b_X ({\btheta}) b_Y ({\btheta}') 
\langle \delta I_{\rm pt}({\bf r}'-L {\btheta}) 
\delta I_{\rm pt}({\bf r}'+{\bf r}-L {\btheta}') \rangle . \label{CXY}
\end{eqnarray}
The Fourier-transformed equivalent of equation (\ref{CXY}) 
illustrates clearly the connection between source structure and the 
scintillation statistics via the power spectrum of intensity 
fluctuations,
\begin{eqnarray}
W_{XY}({\bf q}) = \widetilde{C_{XY}} ({\bf q}) &=& \langle \widetilde{S}_X({\bf q}) 
\widetilde{S}_Y^*({\bf q}) \rangle \nonumber \\
&=& \frac{1}{L^4} [ \widetilde{B}_X({\bf q}) \widetilde{B}_Y^*({\bf q})
{W_{II}}_{\rm pt}({\bf q})],\label{StructEqun}
\end{eqnarray}
where $B_X ({\btheta}) = b_X ({\btheta}/L)$, the 
Fourier transform of a function is denoted with a tilde, and we have 
made use of equation (\ref{CWrelation}). The quantity $B_X(\btheta)$ is
identical to the visibility $\Gamma_X(0;{\bf r})$ of the previous section.
Equation (\ref{StructEqun}) may be inverted to derive information on the 
structure of a scintillating source if the scattering properties of the 
ISM, embodied in ${W_{II}}_{\rm pt}({\bf q})$, are known.  

Equation (\ref{StructEqun}) may be interpreted as follows.  The 
quantity ${W_{II}}_{\rm pt} ({\bf q})$ is the power spectrum of intensity 
fluctuations in the scintillation pattern that would be observed for a 
point source.  The quantity $\widetilde{B}_X({\bf q}) 
\widetilde{B}_Y^*({\bf q})$ is the power spectrum of intensity 
structure in the source.  The resultant power spectrum of intensity 
fluctuations, $\widetilde{C_{XY}}({\bf q})$, is that of a point source 
if the angular structure contained in source is smaller than the scale 
of the scintillation pattern for a point source (ie. the source is 
unresolved by the ISM).  On the other hand, 
the observed scintillation pattern is dominated by source angular 
structure if the source is larger than the scale of the scintillation 
pattern for a point source.  For example, scintillations in the weak 
scattering regime contain power on length scales $r \gtrsim 
r_{\rm F}$ (equivalent to $q \lesssim 1/r_{\rm F}$).  Source structure 
has no measurable effect if the source angular size is less than the angular scale 
of the scintillation pattern (ie.  when $\theta_S < r_{\rm F}/D$).


If the angular size of the source is smaller than the scale size of 
the scintillation pattern for a point source, the term in square 
brackets in (\ref{StructEqun}) is approximated by ${W_{II}}_{\rm pt} ({\bf q})$.  
This is because ${W_{II}}_{\rm pt} ({\bf q})$ decreases much faster 
with $q$ than $\widetilde{B_I}({\bf q}) \widetilde{B_I}^*({\bf q})$.
On the other hand, if the source angular size exceeds that of the 
scintillation pattern for a point source, we can use the approximation
${W_{II}}_{\rm pt} \approx 2.8 \, q^{4-\beta} r_{\rm F}^4/r_{\rm 
diff}^{\beta-2}$ for small $q$ to obtain (see equation (\ref{CXY}))
\begin{eqnarray}
C_{XY}({\bf r}) \approx 
\frac{2.8 \, r_{\rm F}^4 r_{\rm diff}^{2-\beta}}{(2 \pi)^2 L^4} 
\int d^2{\bf q} \, e^{-i {\bf q} \cdot {\bf r}} q^{4-\beta} 
\tilde{B_X}({\bf q}) \tilde{B_Y}^* ({\bf q}) .  
\label{CXYsource}
\end{eqnarray}
Sources sufficiently large to be resolved by the scattering medium 
exhibit smaller and slower intensity fluctuations relative to unresolved sources.
The intensity modulation index, $C_{XY}(0)$, of an resolved source is 
therefore smaller than that of an unresolved source.  This is evident from 
equation (\ref{StructEqun}).  The amplitude of $C_{XY}(0)$ decreases 
if the terms due to the source cut off $C_{XY}({\bf q})$ before 
${W_{II}}_{\rm pt}$ reaches its maximum value (e.g. at 
$q \sim 1/r_{\rm F}$ for weak scattering).

\section{Earth Orbit Synthesis} \label{Combined}

The earth's orbital motion allows the scintillation pattern of a 
scattered source to be mapped in two dimensions.  This is because the 
velocity of an earth-based observer relative to the ISM changes as the 
earth's orbital velocity changes during the course of a year.  An 
observer moving at a velocity ${\bf v}$ relative to the ISM 
samples the scintillation pattern along the direction parallel to 
${\bf v}$.  It is possible to measure the correlations $\Gamma_X({\bf r}={\bf
v}t)$ or $C_{XY}({\bf r}={\bf v} t)$ for observations a time $t$ apart.  
The quantity $\Gamma_X({\bf r})$ can be measured over short coherent
integrations (see, e.g. Goodman \& Narayan 1989).  A good approximation to 
the quantity $C_{XY}({\bf r})$ can typically be derived from observations 
of $\gtrsim 5$ scintles, typically requiring one to two days flux density 
monitoring for typical IDV sources.

Over the course of the year observations at different epochs sample  the
scintillation pattern in different directions, as the velocity of  the earth
changes around the solar circle.  In particular, if the  velocity of the ISM is
comparable to the earth's, then a strong  ``annual cycle'' effect will appear
in the observed IDV pattern, as  has been observed already in J1819+3845
(Dennett-Thorpe \& de Bruyn  2001a) and 0917+624  (Jauncey \& Macquart 2001;
Rickett \et\ 2001).  This enables measurement of $\Gamma_X({\bf r})$ or
$C_{XY}({\bf r})$ for a large range of orientations of ${\bf r}$, and the
information may be inverted to obtain source structure provided that it is 
stable on a timescale long enough for it to be mapped.  The main limitation to  the technique is that one does not always know ${\bf v}$ because the  velocity of the ISM is not generally 
known.  Thus one must solve  simultaneously for ${\bf v}$ and the source 
structure, or measure ${\bf v}$ directly from observations of the time delay between the 
scintillation pattern as observed at two widely separated ($\gtrsim  3000$~km)
receivers as the scintillation pattern passes across the  earth (Jauncey \et\ 2001, 
Dennett-Thorpe \& de Bruyn 2002).

We demonstrate the technique with several examples, concentrating on 
Earth Orbit Synthesis using intensity fluctuations.

\subsection{An example using Intensity Fluctuations}\label{IntensityFlucts}

An intensity image of the structure in a scintillating source requires 
measurement of the autocorrelation function $C_{II}({\bf r})$.  It is not 
possible to measure the spatial structure of the scintillation pattern 
of a source, $C_{II}({\bf r})$, directly.  The scale of the 
scintillation pattern is usually larger than that of earth so that it 
is impossible to measure $C_{II}({\bf r})$ on sufficiently long 
baselines ${\bf r}$.  This problem is overcome by measuring the 
correlations of the scintillation signal as a function of time, rather 
than baseline.  One measures the temporal correlation function
\begin{eqnarray}
    {C_{II}}_t(t) = \frac{\langle I(t')I(t+t') \rangle}{\langle I 
    \rangle^2} -1.
\end{eqnarray}
The correlation between two scintillation signals measured a time $t$ 
apart at the same receiver may be derived from the correlation in the 
scintillation pattern measured by two receivers a distance ${\bf r}$ 
apart.  The relation between the temporal and spatial intensity 
autocorrelation functions is thus
\begin{eqnarray}
    {C_{II}}_t(t)=C_{II}({\bf r}={\bf v} t) = \frac{1}{(2 \pi)^2 \, L^4}
    \int d^2{\bf q} \, e^{-i 
    {\bf q} \cdot {\bf r}} \left[ 
    \widetilde{B_I}({\bf q}) \widetilde{B_I}^*({\bf q})
    {W_{II}}_{\rm pt} ({\bf q}) \right]. \label{CXYr}
\end{eqnarray}
An important point in connection with (\ref{CXYr}) is that the 
modulation depth of the scintillation pattern, $C_{II}(0)$ 
is independent of the velocity of the scattering medium relative to 
earth.  For instance, the depth of 
the intensity modulations does not depend on whether the scintillation 
pattern moves parallel or orthogonal to the long axis of an elongated 
source.  However, we see that the characteristic timescale of the scintillations depends strongly on the direction of motion if ${\bf v}$ varies during the course of a year.

The effect of source structure is particularly simple when a source 
undergoes weak scattering and its angular size is larger than that of 
the scintillation pattern of a point source 
(ie. $\theta_S > r_{\rm F}/D$).  Equation 
(\ref{CXYsource}) is applicable in this case, and we assume $\beta=4$ 
instead of the 
usual Kolmogorov value of $\beta=11/3$ to simplify the algebra and 
derive analytic results.  We obtain
\begin{eqnarray}
C_{II}({\bf r}) = \frac{2.8 \, r_{\rm F}^4}{r_{\rm diff}^2 L^4}
\int d^2{\bf q} \, e^{i {\bf q} \cdot {\bf r}} \, 
| \widetilde{B_I}({\bf q}) |^2, \qquad L \theta_S > r_{\rm F}. 
\label{CIIsimple}
\end{eqnarray}
This expression, and the assumptions made in deriving it, are employed in 
the following examples.

\subsubsection{Elongated source}

We illustrate Earth Orbit Synthesis by examining the combined effect of 
source structure  and velocity variations on a source with elliptical 
brightness distribution:
\begin{eqnarray}
b_I({\btheta}) = \frac{1}{\alpha_x \alpha_y \pi} \exp\left[- 
\left(\frac{\theta_x}{\alpha_x}\right)^2 - 
\left(\frac{\theta_y}{\alpha_y}\right)^2 \right], \label{ElliptSource}
\end{eqnarray}
where $\alpha_x$ and $\alpha_y$ are chosen to be the source angular scales parallel to 
the ${\bf e}_{\parallel}$ and ${\bf e}_{\perp}$ axes respectively.  This choice is 
made for convenience only, and it is simple to re-orient the source at 
an angle $\xi$ to the axes by 
making the transformation $(x,y) \rightarrow (x \cos \xi + y \sin 
\xi, -x \sin \xi + y \cos \xi)$.
Combining equations (\ref{CIIsimple}) and (\ref{ElliptSource}) yields 
the following intensity autocorrelation 
function
\begin{eqnarray}
C_{II}({\bf r}) = \frac{2.8 \, r_{\rm F}^4}{2 \pi \alpha_x \alpha_y 
L^2 r_{\rm diff}^2} \exp \left[- \left( \frac{x^2}{2 L^2 \alpha_x^2}+ 
\frac{y^2}{2 L^2 \alpha_y^2} \right)\right], \qquad L \alpha_{x,y} > 
r_{\rm F} \label{ElliptDecor}
\end{eqnarray}
It is possible to determine the temporal decorrelation of the 
scintillation signal, for a given scintillation speed and direction.  In the 
following we use $V_{\parallel}=15$~km/s, $V_{\perp}=10$~km/s and we 
assume the source has ecliptic co-ordinates $\theta=30^{\circ}$ and 
$\phi=0$.  The variation in the scintillation speed over the course of 
a year is shown in Fig.  3.  Figures 4(a) and 5(a) show how the 
intensity temporal autocorrelation function $C_{II}(t)$ varies 
throughout the year for an elliptical source with these screen 
parameters.

The effect of source structure is evident in the timescale of 
variability.  It is convenient to define the decorrelation timescale, 
$t_{\rm dc}$ as the time in which the intensity correlation falls to 
$1/e$ of its maximum value.  The decorrelation timescale for the 
elliptical source described by equation (\ref{ElliptDecor}) is
\begin{eqnarray}
t_{\rm dc} = \left[ \frac{({\bf v}_{\rm scint} \cdot {\bf 
e}_{\parallel})^2}{2 L^2 \alpha_x^2} + \frac{({\bf v}_{\rm scint} 
\cdot {\bf e}_{\perp})^2}{2 L^2 \alpha_y^2} \right]^{-1/2}.
\end{eqnarray}
Changes in the magnitude of ${\bf v}_{\rm scint}$ cause changes in the 
scintillation timescale even if the source is circularly symmetric, 
whereas changes in the direction of ${\bf v}_{\rm scint}$ only result in changes 
in $t_{\rm ISS}$ if the source is elongated, when $\alpha_x \neq 
\alpha_y$.

Figures 4(b) and 5(b) plot the variation in decorrelation timescale as 
a function of epoch for a specific scintillation velocity for the 
elliptical source with two different source parameters, $L \, \alpha_x 
=10^{9}$~m and $L\, \alpha_y = 10^{9}$~m and $5\times 10^{9}$~m 
respectively.  Fig. 4(b) shows the expected 
scintillation timescale for a circularly symmetric source, and Fig. 5(b) 
the timescale for an elongated source.  The peak in the 
scintillation timescale at $T \approx 0.9$~yr is common to both plots 
and occurs because the scintillation speed reaches a minimum.  The 
emergence of another peak in Fig.  5(b) occurs due to the presence of 
structure in the source, here an elongation along the $y$-axis.  The 
peak occurs at the time of the year that the scintillation direction 
is tangential to the long axis of the source.

\subsubsection{Two-component source}

A source with two components shows richer structure in the statistics 
of the intensity variations.  We 
consider two gaussian patches of emission, the central peak having an 
intensity $a_0 I_0$ ($0 < a_0 < 1$) and the other, located at an angular 
separation of ${\btheta}_0$, having an intensity $(1-a_0) I_0$.   
The two subsources have identical angular scales $\alpha$ and we again assume that the 
angular diameters of the subsources exceed the angular diameter of 
the scintillation pattern due to a point source, so that equation 
(\ref{CXYsource}) applies in the regime of weak scattering. 
The source angular brightness distribution is 
\begin{eqnarray}
b_I({\btheta}) = \frac{a_0}{\pi \alpha^2} \exp\left[- 
\frac{{\btheta}^2}{\alpha^2}  \right] + 
\frac{1-a_0}{\pi \alpha^2} \exp\left[- \frac{({\btheta} 
-{\btheta}_0)^2}{\alpha^2} \right],
\label{TwoSource}
\end{eqnarray}
for which the resulting spatial intensity autocorrelation function is 
\begin{eqnarray}
C_{II}({\bf r}) =  \frac{r_{\rm F}^4}{4 \alpha^2 L^2 \pi^2 r_{\rm diff}^2}  
\exp \left[ - \frac{r^2+ L^2 {\btheta}_0^2}{2 \alpha^2 L^2} \right] 
\left( (1 - 2 a_0 + 2 a_0^2) \exp\left[ \frac{ {\btheta}_0^2}{2 \alpha^2 }\right]
+2 a_0 (1-a_0) \cosh \left[\frac{ {\bf r} \cdot {\btheta}_0}{\alpha^2 L} \right] \right).
\label{TwoSourceC}
\end{eqnarray}
Figure 6 plots the spatial intensity autocorrelation function of this source.  This 
function has three peaks.  The lightcurve of a two-component source 
therefore exhibits interesting behaviour when the scintillation direction is 
parallel to the line joining the two substructures.  An observer then 
measures the spatial autocorrelation function along a line that 
joins the three peaks in the 
autocorrelation function.

A specific example of such behaviour is illustrated in Figure 7, which 
demonstrates how the temporal autocorrelation function of a two 
component source changes throughout the course of a year.  At 
$T\approx 0.0-0.2$~yr the scintillation velocity is parallel to the 
direction of the two separated peaks, and a second bump is visible in 
the temporal intensity autocorrelation function.

\subsubsection{Polarization reference mapping}

Only limited information about source structure is available from Earth 
Revolution Synthesis using intensity fluctuations.  It is impossible 
to determine $b_I(\btheta)$ directly from $C_{II}({\bf r})$.  Only the correlation 
of the angular brightness distribution, $\langle b_I(\btheta +\btheta') 
b_I(\btheta')\rangle$, can be derived.  

However, Earth Orbit Synthesis is useful even when the brightness distribution 
$b_I ({\btheta})$ cannot be ascertained exactly.
IDV sources often exhibit fluctuations in Stokes $Q$, 
$U$ and $V$ that are dissimilar to those observed in $I$.  This occurs if 
the polarized structure of the source differs from its structure in 
total intensity.  This substructure can be mapped relative to the 
total intensity structure, using the correlations $C_{IQ}({\bf r})$, 
$C_{IU}({\bf r})$ and $C_{IV}({\bf r})$ to find the quantities
$\langle b_I ({\btheta'+\btheta}) b_Q ({\btheta'})\rangle$, etc..
It is thus possible to map the brightness distribution 
of a source in one Stokes parameter relative to another Stokes 
parameter, as has been shown for both the linear and circular polarized
components in PKS~1519$-$273 (Macquart \et\ 2000).

\section{Conclusion} \label{Concl}

Measurement of the scintillation characteristics of IDV sources allows their
structure to be investigated with microarcsecond resolution.  The radio 
structures of IDV sources are currently of interest because of their  high
brightness temperatures.  Earth Orbit Synthesis has the potential to determine
the origin of these compact structures; it can determine, for example, whether
the observed emission emanates from the central cores of these objects or from
the bases of jets. 

Short timescale polarization variability in IDV sources indicates the presence
of polarization structure on similarly fine scales and --- in many cases --- on
even smaller scales than the structure in total intensity.  The advantage of
Earth Orbit Synthesis over traditional polarimetric imaging is its high
resolution.  Such a high resolution probe is less subject to
beam-depolarization.  Indeed, this mapping technique is already discovering
remarkable detail about the polarized sub-structure of IDV sources (e.g.
Macquart \et\ 2000).

The observed two-dimensional scintillation pattern of a scattered radio source 
is highly influenced by source structure.  This 
pattern can be measured in two dimensions if the apparent velocity of 
the scintillation across the line of sight to a scattered source 
changes; changes in direction allow an observer to probe source 
structure along different directions.  Direction changes occur when the 
velocity of the scattering material is comparable to the earth's 
velocity, and the vector sum of these two velocities changes during 
the course of a year.

The derivation of source structure requires measurement of the 
scintillation fluctuations at several epochs over the course of a 
year.  This is achieved by measuring the temporal decorrelation of the 
intensity $\langle I (t) I(t+t') \rangle$.  Many IDV sources also 
exhibit polarization variability, which is often not well 
correlated with the total intensity.  This occurs when the 
source structure in polarized and total intensity emission differs.  
Polarized substructure information is available by measuring the 
temporal cross-correlations $\langle X(t) Y(t+t') \rangle$ between 
Stokes parameters $X$ and $Y$.  It is possible to measure the 
polarized substructure of a source with reference to the intensity 
emission to high precision.

An advantage of the Earth Orbit Synthesis technique is the capability to
compare source structure between frequencies.  For example, one could compare
the angular separation between source components as a function to yield
information relating to opacity changes with frequency.  Analysis of the data on
PKS~1519$-$273 (Macquart \et\ 2000) demonstrates not only the power to compare
source structure at two frequencies, but even to compare the relative source
properties in different Stokes parameters.

It should be emphasised that the use of Earth Orbit Synthesis as an imaging
technique is subject to limitations.  Firstly, the scintillation process only
probes source structure over a limited angular size range.  Consequently, while
Earth Orbit Synthesis probes microarcsecond structures, it provides no
information on the relation between microarcsecond and larger scale structure. 
Secondly, the use of autocorrelation functions in the imaging process is a
necessary but severe limitation to any imaging that can be achieved.  This
technique yields the power spectrum of the source brightness distribution and,
as such discards phase information that is present in normal rotational
synthesis.  The stochastic nature of the scintillation process forces us to use
ensemble average quantities such as autocorrelation functions.

The angular resolution achievable using scintillation-based imaging 
depends on the properties of the scattering medium.  In the regime of 
weak scattering, generally applicable at frequencies $\nu \gtrsim 5$~GHz, the 
intensity scintillations are sensitive to angular scales 
$\theta \sim r_{\rm F}/D \propto \nu^{-1/2}$.  The scattering is strong at 
frequencies $\nu \lesssim 5$~GHz.  The refractive scintillation 
exhibited by IDV sources in this regime is less sensitive to small 
angular scales because the angular resolution achievable is $\theta 
\sim r_{\rm ref}/D$ which is large, and scales as $\nu^{-2.2}$ for a 
Kolomogorov spectrum of inhomogeneities.  Strong scattering is typically
sensitive to sizes $\gtrsim 0.1$~mas, and hence can be found in many more
flat-spectrum sources, particularly at low Galactic latitudes.  Thus, while
less useful competing against VLBI for milliarcsecond imaging, strong scattering here
remains an excellent probe of the ISM close to the Galactic plane.

A program to use Earth Orbit Synthesis to map structure in IDV sources 
is currently underway using the Australia Telescope Compact Array (Bignall \et\
2001).

\acknowledgments
 
The authors thank Barney Rickett, Simon Johnston and Lucyna Kedziora-Chudczer 
for valuable discussions and comments on the manuscript.  JPM is particularly 
indebted to Don Melrose, from whose Festschrift the idea for this paper originated.
The Australia Telescope is funded by the Commonwealth Government for operation
as a national facility by the CSIRO.

\clearpage

\begin{figure}
\plotone{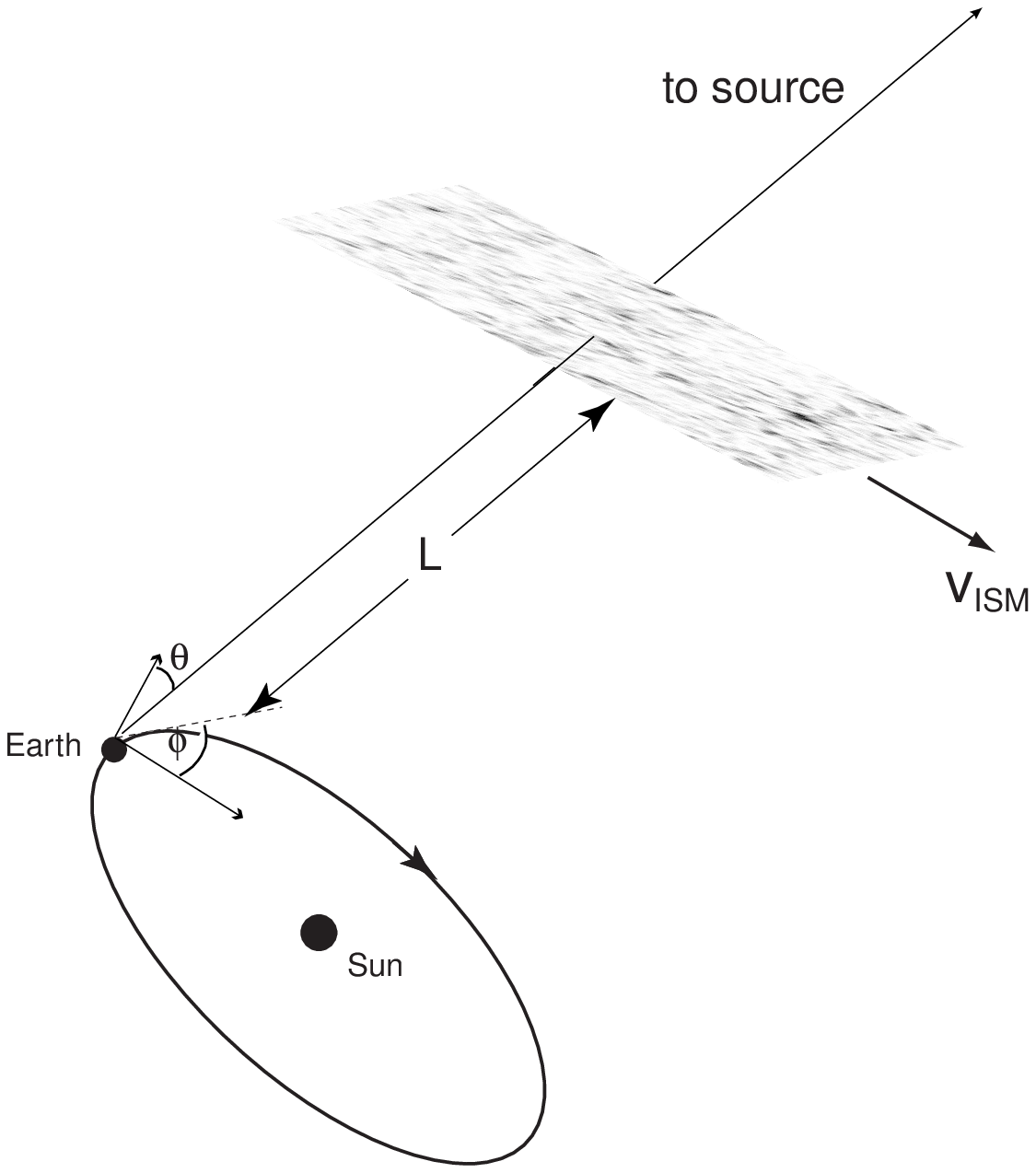}
\caption{The scattering model.  The earth-source line of sight 
moves relative to a scattering screen, fixed at a distance $L$ from 
earth.}
\end{figure}

\begin{figure}
\plotone{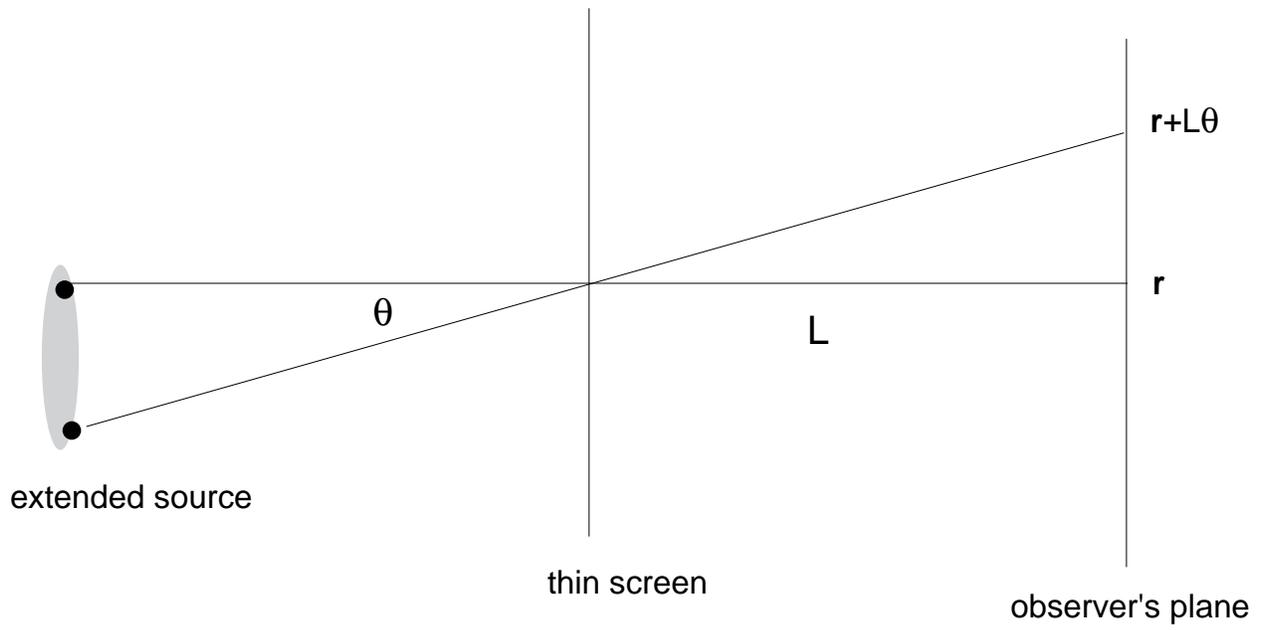}
\caption{The geometry of the scattering assumed in calculating the 
effect of a source's structure on its scintillation pattern.}
\end{figure}

\begin{figure}
\plotone{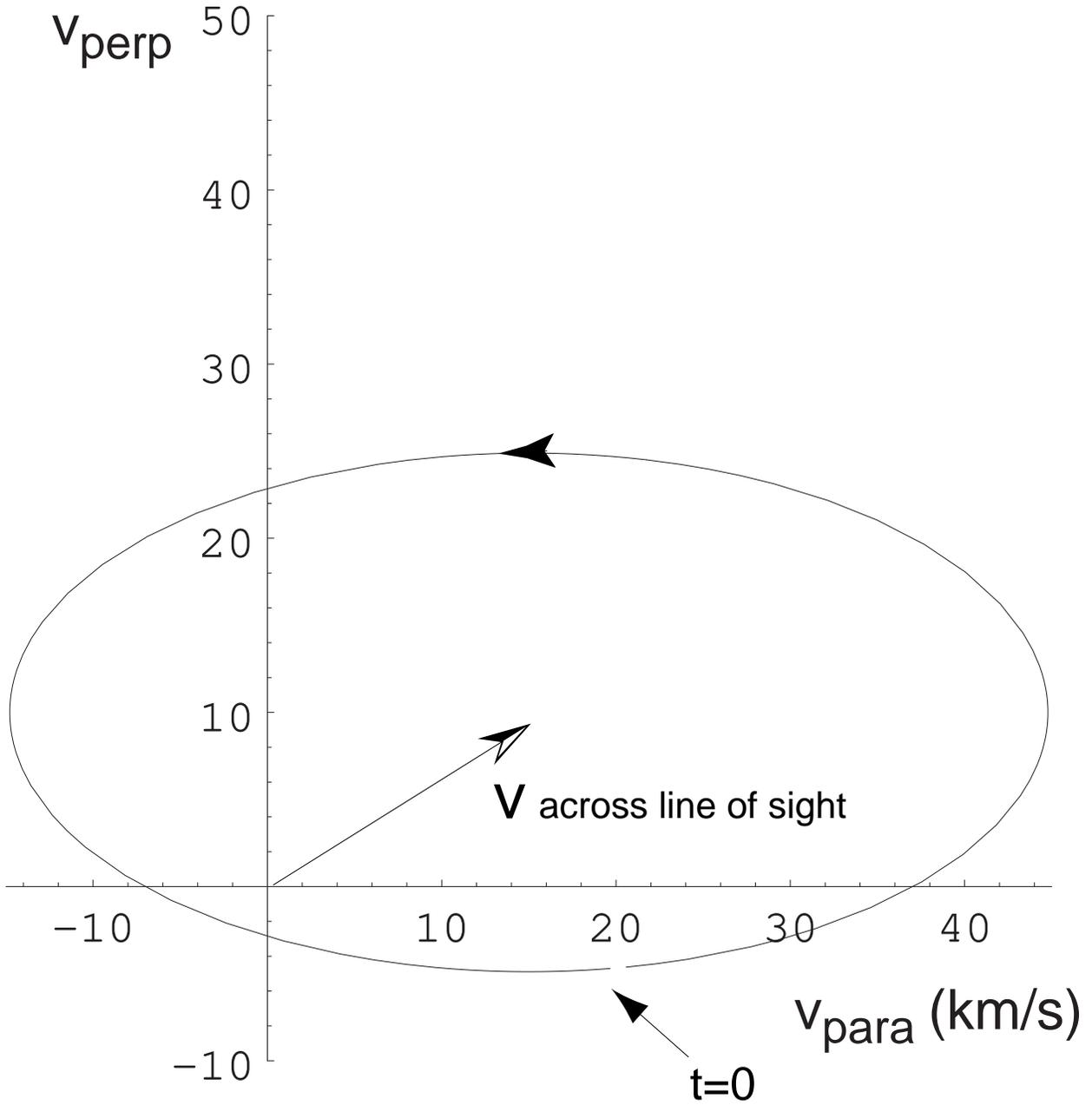}
\caption{The scintillation speed of a source located at ecliptic 
co-ordinates $(\theta,\phi)=(30^{\circ},0^{\circ})$ with an intrinsic 
ISM velocity of $(V_{\parallel},V_{\perp})= (15,10)$~km/s.}
\end{figure}

\begin{figure}
\plottwo{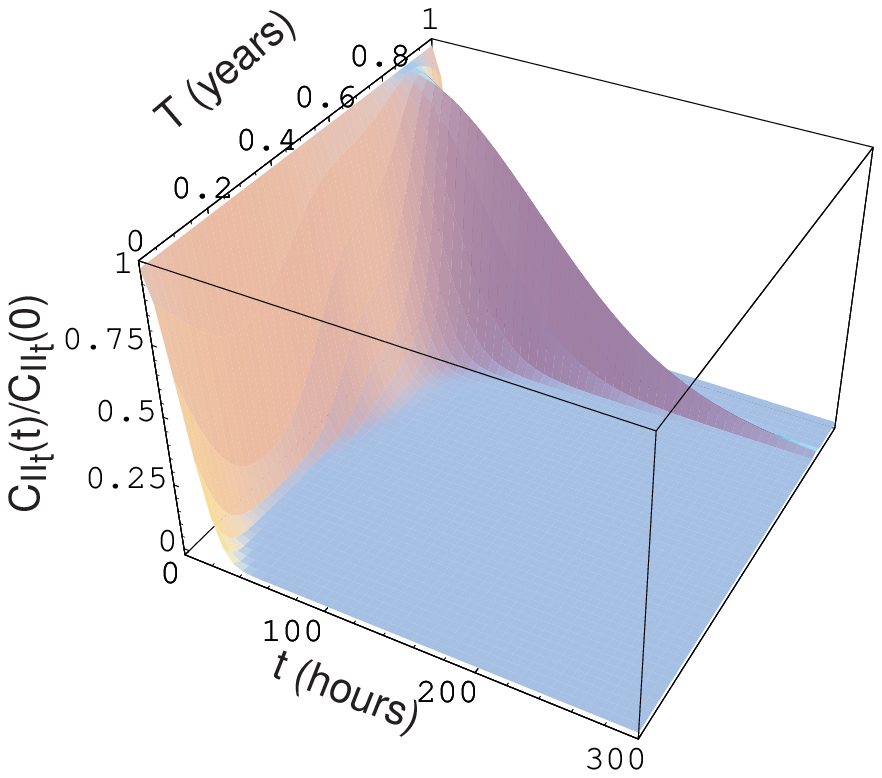}{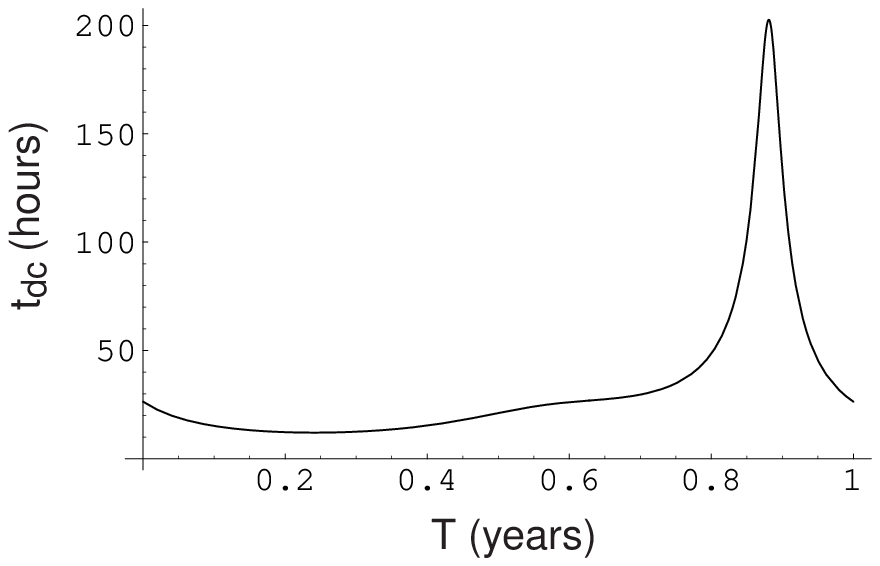}
\caption{The observed intensity autocorrelation function as a function of  time
and a epoch for an elliptical source  whose angular size exceeds scattering
disk and with a  brightness distribution given by equation (\ref{ElliptSource})
with  $L \, \alpha_x = 10^9$~m and $L \, \alpha_y=10^9$~m.  Panel (a) shows the
evolution  of the temporal autocorellation function, ${C_{II}}_t$ throughout
the  year in the limit $\beta \rightarrow 4$, while (b) shows the evolution of
the decorrelation timescale.}
\end{figure}

\begin{figure}
\plottwo{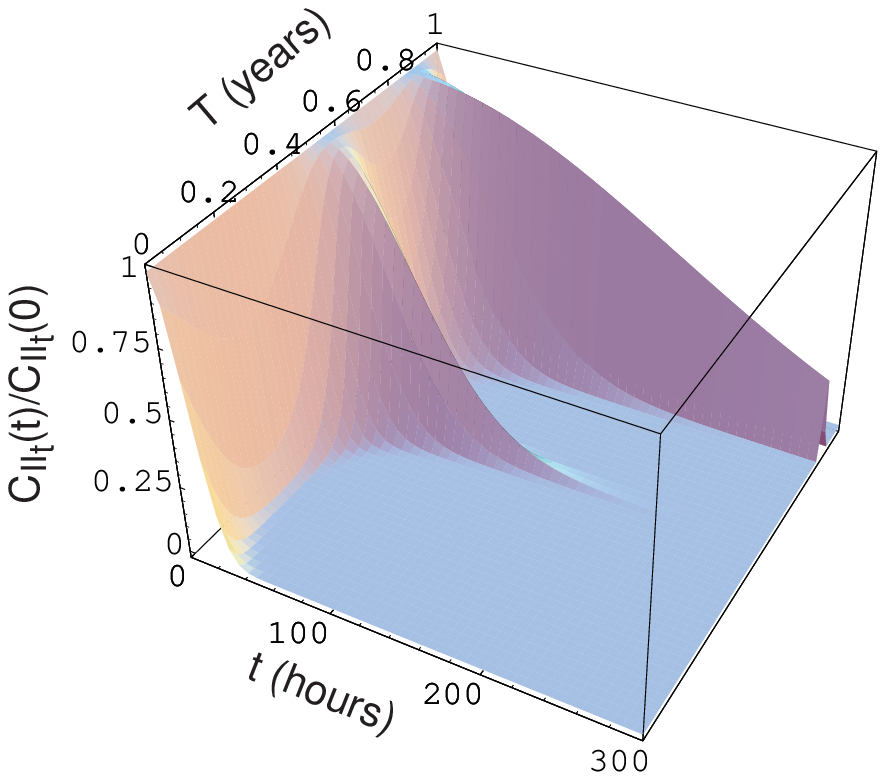}{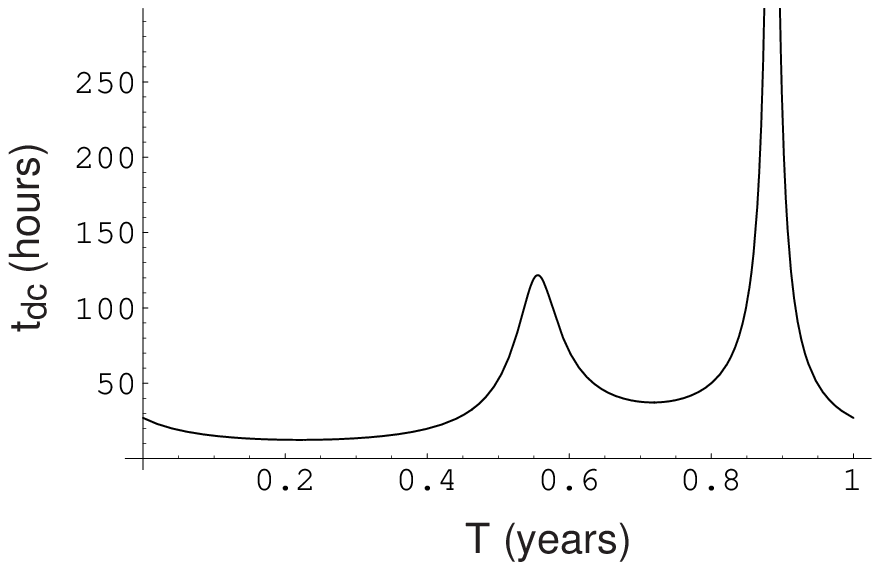}
\caption{The observed intensity autocorrelation function as a function of  time
and a epoch for a source whose angular size exceeds scattering disk and with a 
brightness distribution given by equation (\ref{ElliptSource}) with  $L \,
\alpha_x = 10^9$~m and $L \, \alpha_y=5 \times 10^9$~m.  Panel (a) shows the
evolution  of the temporal autocorellation function, ${C_{II}}_t$ throughout
the  year (in the limit $\beta \rightarrow 4$), while (b) shows the evolution
of the decorrelation timescale.}
\end{figure}

\begin{figure}
\plotone{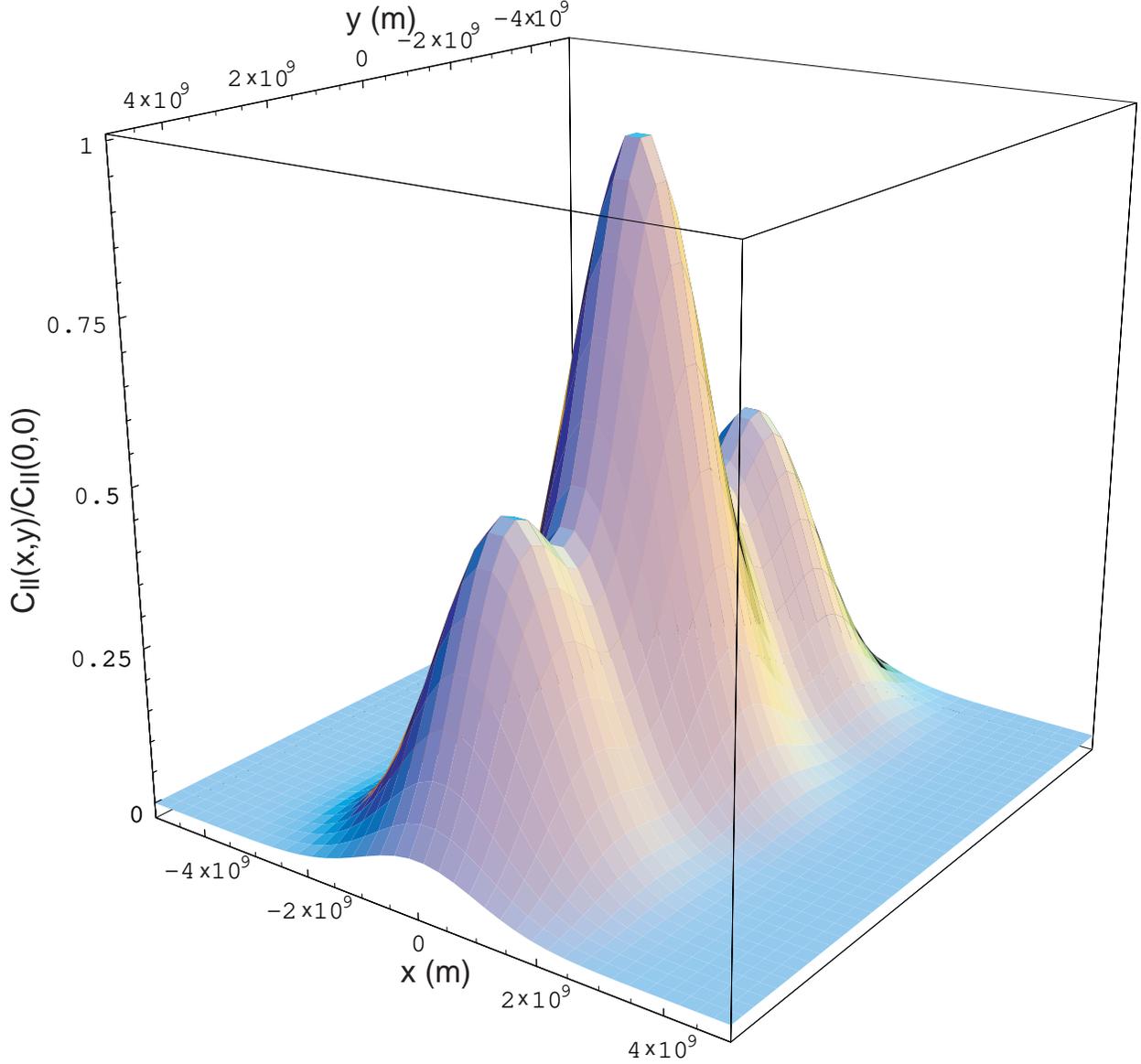}
\caption{The spatial intensity autocorrelation function of a source 
whose angular size exceeds that of the scattering disk and with a 
brightness distribution given by equation (\ref{TwoSource}) with 
$L \alpha=10^9$~m and a source separation of 
$L ({\theta_0}_x,{\theta_0}_y)=(3,0) \times 10^9$~m, $a_0=0.5$ and 
assuming $\beta \rightarrow 4$.}
\end{figure}

\begin{figure}
\plottwo{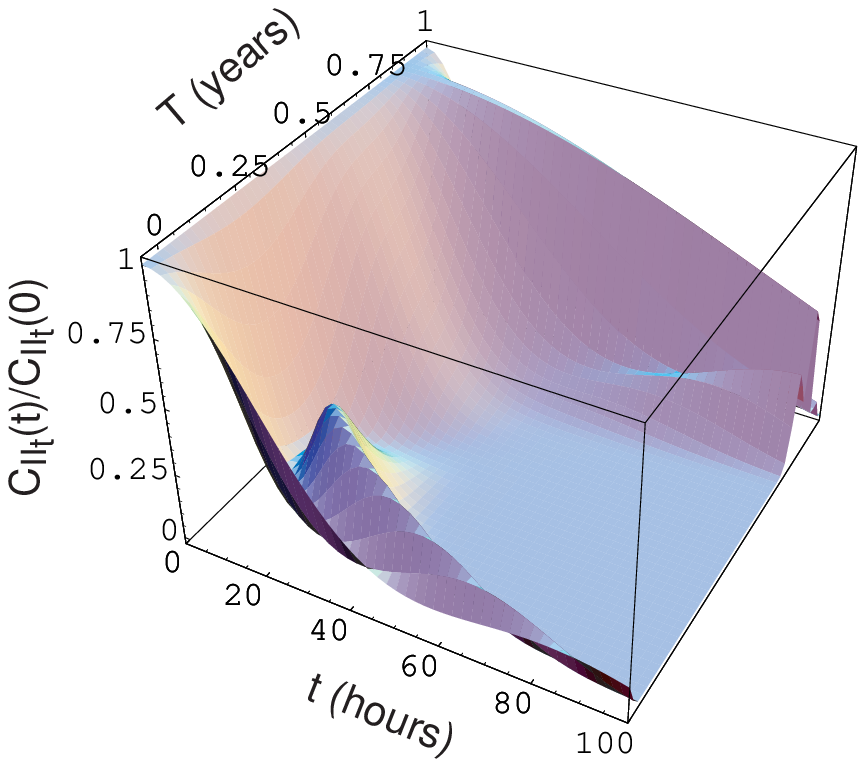}{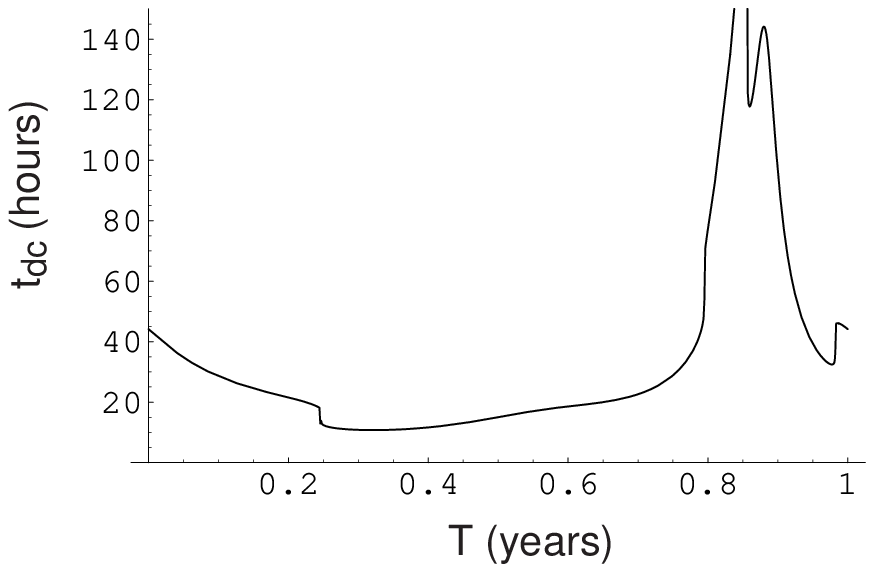}
\caption{The observed intensity autocorrelation function as a function of 
time and a epoch for a source whose angular size exceeds scattering disk and with a 
brightness distribution given by equation (\ref{ElliptSource}) with 
$\alpha_x = 4 \times 10^9$, $\alpha_y=10^9$ and $a_0=0.5$.  Panel (a) shows the evolution 
of the temporal autocorellation function, ${C_{II}}_t$ throughout the 
year ($\beta \rightarrow 4$), while (b) shows the evolution of the 
decorrelation timescale.}
\end{figure}

\end{document}